\documentclass[english,twocolumn, prl, superscriptaddress, longbibliography]{revtex4-1}
\usepackage[T1]{fontenc}
\usepackage{color}
\usepackage{amsmath}
\usepackage{graphicx}
\usepackage{dsfont}
\usepackage[colorlinks=true,linkcolor=blue,urlcolor=blue,citecolor=blue,anchorcolor=blue]{hyperref}
\makeatletter
\usepackage{amsmath}
\usepackage{subfigure}
\usepackage{graphicx, mathtools}
\usepackage[normalem]{ulem}

\makeatother

\usepackage{babel}

\begin{document}

\newcommand{\be}{\begin{equation}}
\newcommand{\ee}{\end{equation}}
\newcommand{\bn}{\begin{eqnarray}}
\newcommand{\en}{\end{eqnarray}}
\newcommand{\ii}{\'{\i}}
\newcommand{\ca}{\c c\~a}
\newcommand{\uc}{\uppercase}

\title{Quantum Critical Magneto-transport at a Continuous Metal-Insulator Transition}

\author{P. Haldar}\email{prosenjit@imsc.res.in}

\author{M. S. Laad}\email{mslaad@imsc.res.in}

\author{S. R. Hassan}\email{shassan@imsc.res.in}

\affiliation{Institute of Mathematical Sciences, Taramani, Chennai 600113, India and \\
Homi Bhabha National Institute Training School Complex,
Anushakti Nagar, Mumbai 400085, India}

\author{Madhavi Chand}\email{madhavichand2009@gmail.com}

\author{Pratap Raychaudhuri}\email{praychaudhuri2014@gmail.com}

\affiliation{Tata Institute of Fundamental Research, Homi Bhabha Rd, Colaba, Mumbai 400 005, India}

\date{\rm\today}

\begin{abstract}
In contrast to the seminal weak localization prediction of a non-critical Hall constant ($R_{H}$) at the Anderson metal-insulator transition (MIT), $R_{H}$ in quite a few real disordered systems exhibits both, a strong $T$-dependence and critical scaling near their MIT.  Here, we investigate these issues in detail within a non-perturbative ``strong localization'' regime using cluster-dynamical mean field theory (CDMFT).  We uncover $(i)$ clear and unconventional quantum-critical scaling of the Gell-Mann-Law, or $\gamma$-function for magneto-transport, finding that $\gamma(g_{xy})=\frac{d[log(g_{xy})]}{d[log(T)]}\simeq$ log$(g_{xy})$ over a wide range spanning the continuous MIT, very similar to that seen for the longitudinal conductivity, $(ii)$ strongly $T$-dependent and clear quantum critical scaling in both transverse conductivity and $R_{H}$ at the MIT.  We show that these surprising results are in comprehensive and very good accord with signatures of a novel Mott-like localization in NbN near the MIT, providing substantial support for our ``strong'' localization view. 
\end{abstract}


\maketitle

  At low temperatures ($T$), transport in normal metals arises as a result of scattering of weakly interacting fermionic (Landau) quasiparticles amongst themselves, phonons 
and impurities~\cite{Nozieres}.  Remarkably, it seems to hold even for $f$-electron systems, which are certainly strongly correlated Fermi liquids.  However, this appealing quasiclassical description fails
near metal-insulator transitions (MITs), where the Landau quasiparticle description itself breaks down~\cite{imada}.  In fact, in 
cuprates~\cite{ong} and some $f$-electron systems~\cite{steglichNature}, resistivity and Hall data can only be reconciled by postulating 
{\it two} distinct relaxation rates, arising from break-up of an electron, for 
the decay of longitudinal and transverse currents.  In many cases, bad-metallic and linear-in-$T$ resistivities preclude use of 
Boltzmann transport views altogether, since the picture of weakly interacting Landau quasiparticles itself breaks down.

  In disorder-driven MITs, resistivity and Hall effect have long been studied in the context of the seminal weak-localization (WL) 
  theory~\cite{paleetvr1985}.  These studies already threw up interesting hints regarding the inadequacy of WL approach upon attempts to 
  reconcile criticality in (magneto)-transport~\cite{rosenbaum}. 
  Specifically, while both $\sigma_{xx}(n)\simeq (n_{c}-n)^{\nu}$ and $\sigma_{xy}(n)\simeq (n_{c}-n)^{\nu'}$ turned out to be critical at the MIT, 
  and the ratio $\nu'/\nu\simeq 1$ in stark contrast to the value of $2$ predicted at the Anderson MIT~\cite{abrahams}.  More recent work on 
  intentionally disordered NbN~\cite{madhavi}, 
  wherein the system is driven across $k_{F}l\simeq O(1)$ (here, $k_{F}$ is the Fermi wave-vector and $l$ the mean-free path, and $k_{F}l>>1$ describes a good metal, while $k_{F}l\leq 1.0$ describes a bad metal without well-defined electronic quasiparticles), shows clear signatures of an unusual type of localization at odds with theoretical 
  predictions if one insists on an Anderson disorder-driven MIT: $(i)$ $\rho_{xx}(T)\simeq C + AR_{H}(T)$, both increasing with reduction in $T$ over a wide range of $k_{F}l$ far {\it before}
  the MIT occurs (in NbN, this is pre-empted by a superconductor-insulator transition (SIT)~\cite{madhavi} at very low $T$ near the critical 
$(k_{F}l)_{c}$, and $(ii)$ $\Delta R_{H}/R_{H} \simeq 0.69(\Delta \rho_{xx}(T)/\rho_{xx})$, widely different from 
$\Delta R_{H}/R_{H} \simeq 2.0(\Delta \rho_{xx}(T)/\rho_{xx})$ expected to hold in WL theory~\cite{Altshuler} ($k_{F}l>>1$).  These
anomalies in both $\rho_{xx}$ and $R_{H}(T)$ are inexplicable within WL views 
(where $R_{H}$ is $T$-independent and {\it non-critical} at the MIT), and point toward a fundamentally new mechanism at work.  Two
possible reasons for this discord are: $(1)$ electron-electron ($e-e$) interactions grow~\cite{Punnoose289} near a disorder-induced MIT and may destroy the one-electron picture, and/or $(2)$ such experiments maybe probing the ``strong'' localization regime of a disorder problem, where non-perturbative strong scattering effects may also destroy the one-electron picture.  
This is because Boltzmann approaches are untenable at the outset when $k_{F}l\simeq 1$, when a quasiparticle
view itself breaks down.  

  Motivated by the above issues, we investigate magneto-transport near a {\it continuous} (at $T=0$) MIT.  
  We choose the 
Falicov-Kimball model because $(i)$ it is the simplest model of correlated fermions exhibiting a continuous MIT, $(ii)$ is {\it exactly} soluble within (cluster) 
dynamical mean-field theory ((C)DMFT) for arbitrarily strong interaction, and $(iii)$ a two-site cluster-DMFT treats the all-important short-range correlations 
precisely on the length scale of $l\simeq k_{F}^{-1}$.  Moreover, it is isomorphic to the binary-alloy Anderson disorder model, except that the FKM has {\it annealed} instead of quenched disorder.  The Hamiltonian is

\be
H_{FKM}= -t\sum_{<i,j>}(c_{i}^{\dag}c_{j} + h.c) + U\sum_{i}n_{i,d}n_{i,c} + \mu \sum_{i}n_{i,c}
\label{equ:equ1}
\ee      

on a Bethe lattice with a semicircular band density of states (DOS) as an approximation to a $D=3$ lattice. 
\begin{figure} 
\includegraphics[width=1.0\columnwidth , height=
1.0\columnwidth]{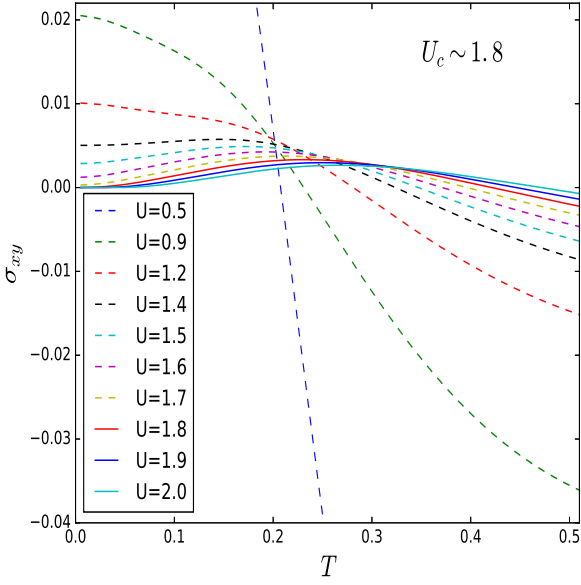}
\caption{(Color online) Hall Conductivity($\sigma_{xy}$) as a function of temperature(T) for different U}
\label{fig:fig1}
\end{figure} 
$c_{i}(c_{i}^{\dag}),d_{i}(d_{i}^{\dag})$ are fermion operators in dispersive band ($c$) 
and dispersion less ($d$) states, $t$ is the one-electron hopping integral and $U$ is the onsite repulsion for a site-local doubly occupied 
configuration.  Since $n_{i,d}=0,1$, $v_{i}=Un_{i,d}$ is also viewed as a static but {\it annealed} ``disorder'' potential for the $c$-fermions. We take non-interacting $c$-fermions half-bandwidth as unity i.e. $2t=1$.
\begin{figure}
\includegraphics[width=1.\columnwidth , height= 
1.\columnwidth]{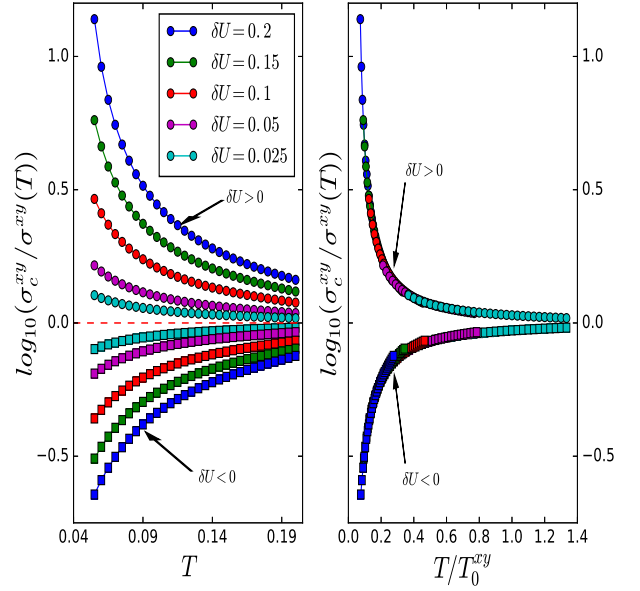} 
\caption{(Color online)(a) In left panel, $log_{10}(\frac{\sigma^{xy}_{c}}{\sigma^{xy}(T)})$ as a function of temperature T for $\delta U=\pm0.025,0.05,0.1,0.15,0.2$; $\sigma^{xy}_{c}$ is the "separatrix". (b)In right panel, scaling the data along T-axis by scaled temperature $T^{xy}_0$.}
\label{fig:fig2}
\end{figure}

    As studied earlier for the dc resistivity~\cite{qcmott}, we now use the exact-to-$O(1/D)$ cluster propagators $G_{\bf K}(\omega)$ for each of the $2$-site cluster 
   momenta ${\bf K}=(0,0),(\pi,\pi)$ to compute the full conductivity tensor, $\sigma_{ab}(T)$, with $a,b=x,y$.
   We neglect the vertex corrections to the Bethe-Salpeter equation (BSE) for all the intra-cluster momenta
   since they are negligible even within CDMFT, as can be seen, for example, within a cluster-to-orbital 
mapping~\cite{silke,kotliar}.  Thus, this constitutes an excellent approximation for computation of transport coefficients.  Explicitly, the dc conductivity reads

\be
\sigma_{xx}(T)=\sigma_{0}\sum_{\bf K}\int_{-\infty}^{+\infty} d\epsilon v^2(\epsilon) \rho_{0}^{\bf K}(\epsilon) \int_{-\infty}^{+\infty} d\omega A_{\bf K}^{2}(\epsilon,\omega)(\frac{-df}{d\omega})
\label{equ:equ2}
\ee
where $\sigma_{0}=\frac{\pi e^{2}}{\hbar Da}\simeq (10^{-3}-10^{-2})(2/D) (\mu\Omega)$.cm${-1}$, $\rho_{0}^{\bf K}(E)$ the ``partial'' 
unperturbed DOS used in earlier work~\cite{ourfirstpaper} and $A_{\bf K}(E)$ the intra-cluster CDMFT one-fermion spectral function. 
The Hall conductivity is a more delicate quantity to compute~\cite{lange}.  Fortunately,absence of vertex corrections comes to the rescue
and we find

\be
\sigma_{xy}(T)=\sigma_{xy,0}B\sum_{\bf K}\int d\epsilon v^2(\epsilon) \rho_{0}^{\bf K}(\epsilon) \epsilon \int d\omega A_{\bf K}^{3}(\epsilon,\omega)(\frac{df}{d\omega})
\label{equ:equ3}
\ee
with $\sigma_{xy,0}=-\frac{2\pi^{2}|e|^{3}a}{3\hbar^{2}}(1/2D^{2})$, and $B$ the magnetic field.  Now, the Hall constant is simply $R_{H}(T)=\frac{\sigma_{xy}}{B\sigma_{xx}^{2}}$ and the Hall angle is
cot$\theta_{H}=\frac{\sigma_{xx}}{\sigma_{xy}}$.
%\begin{figure}
%\includegraphics[width=1.1\columnwidth , height= 
%1.1\columnwidth]{fig1.eps} 
%\caption{(Color online) Hall Conductivity($\sigma_{xy}$) as a function of temperature(T) for different U.}
%\label{fig:fig1}
%\end{figure}
   we show the off-diagonal conductivity, $\sigma_{xy}(U,T)$ as a function of $U$ from small- to large $U$ across the 
  continuous MIT occurring at $U_{c}=1.8$~\cite{ourfirstpaper}.  First, we show results for the temperature-dependent
off-diagonal conductivity, $\sigma_{xy}(T)$ as a function of $U$ across the
continuous Mott transition.

\begin{figure}
\includegraphics[width=1.\columnwidth , height= 
1.\columnwidth]{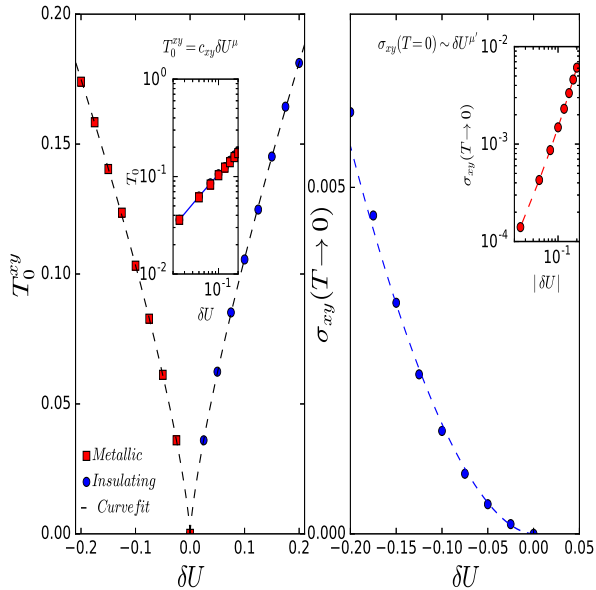} 
\caption{(Color online) (a) In left panel, scaling parameter $T^{xy}_{0}$ as a function of control parameter $\delta U=U-U_c$; the inset 
illustrates power law dependence of scaling parameter $T^{xy}_{0}=c\mid \delta U \mid^{\mu}$. (b)In right panel, $\sigma_{xy}(T \rightarrow 0)$
as a function of control parameter $\delta U=U-U_c$; the inset illustrates power law dependence of
$\sigma_{xy}(T\rightarrow 0)=c\mid \delta U \mid^{\mu'}$.}
\label{fig:fig4}
\end{figure}

   We use Equation~\ref{equ:equ3} to compute $\sigma_{xy}(T,U)$. In Fig.~\ref{fig:fig1} we show  $\sigma_{xy}(T,U)$ as a function of temperature (T) for different disorder values (U). A clear change of slope at low $T < 0.05t$ occurs around $U\simeq 1.3$, which seems to
correlate with the bad-metal-to-bad-insulator crossover in the $dc$ resistivity
in our earlier study~\cite{qcmott}.  Close to the MIT, $\rho_{dc}(T)$ diverges approximately like exp$(E_{g}/k_{B}T)$ as $T\rightarrow 0$ in this regime, $R_{H}(T\rightarrow 0)$ diverges as it must, since the MIT is accompanied by loss of carriers due to gap opening.
A clear change of slope (for $T<0.05t$) occurs around $U=1.3$, and 
  $\sigma_{xy}(T)\simeq T^{1.2}$ around
$U_{c}$.  The $dc$ resistivity $\rho_{xx}(T)$ shows 
extremely bad-metallic behaviour at lowest $T$, beautiful mirror symmetry and 
novel ``Mott-like'' scaling~\cite{qcmott} precisely in this regime.  It is obviously of interest to inquire whether the novel features 
seen in $\rho_{xx}(U,T)$ are also reflected in magneto-transport near 
the ``Mott'' QCP.  To facilitate this possibility, we show $log_{10}(\frac{\sigma^{xy}_{c}}{\sigma^{xy}(T)})$ versus $T$ in the left panel of Fig.\ref{fig:fig2}, finding that the family of $1/\sigma^{xy}(U,T)$ curves also exhibit a near-perfect 
``mirror'' symmetry over an extended region around $1/\sigma^{xy}_{(c)}(U,T)$, strongly presaging quantum critical behaviour.  To unearth this 
feature, we also show $log_{10}(\frac{\sigma^{xy}_{c}}{\sigma^{xy}(T)})$ 
versus $T/T_{0}^{xy}$ in the right panel of Fig.\ref{fig:fig2}, where we have repeated the unbiased method of introducing a $T_{0}^{xy}(U)$ 
to rescale all metallic and insulating curves on to two 
universal curves.  Remarkably, as for the $\rho_{xx}$-scaling, we find, as shown in the left panel of Fig.\ref{fig:fig4}, that $T_{0}^{xy}$ vanishes precisely 
at the MIT.  Clear scaling behaviour we find testifies to a 
remarkable fact: the novel scaling features found earlier in $dc$ resistivity are also clearly manifest in the off-diagonal resistivity.  

  Even clearer characterization of the quantum critical features obtains when we compute the $\gamma$-function~\cite{abrahams} (this is the analogue of the well-known $\beta$-function for the longitudinal conductivity) for $\sigma_{xy}(U,T)$,
  defined by $\gamma(g_{xy})=\frac{d[ln(g_{xy})]}{d[ln(T)]}$, with $g_{xy}=\sigma^{xy}(T)/\sigma^{xy}_{c}$. 
\begin{figure}
\includegraphics[width=1.\columnwidth , height= 
1.\columnwidth]{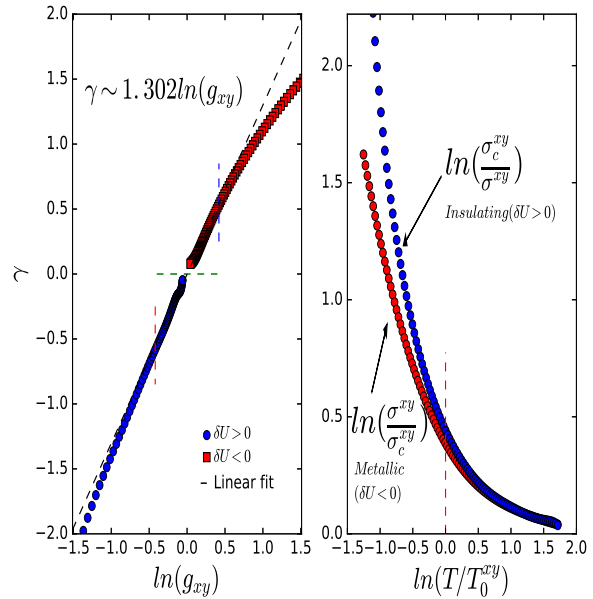} 
\caption{(Color online)(a) In left panel, $\gamma$-function shows linear in $ln(g_{xy})$ behaviour close to the transition. 
Squares are for metallic branch $(\delta U < 0)$ and circles ones are for insulating branch $(\delta U >0)$ ; 
vertical dashed indicate the region where mirror symmetry of curved is found. (b) In right panel, reflection symmetry of scaled curved close to 
the transition.}
\label{fig:fig3}
\end{figure}  
As shown in Fig.\ref{fig:fig3}, it is indeed remarkable that it clearly varies as ln$(g_{xy})$, and is continuous through $\delta U=0$.  This 
shows that it has precisely the same form on both sides of the MIT, which is exactly the feature needed for genuine quantum criticality.  
These features resemble those found for QC scaling in $\rho_{xx}$~\cite{qcmott}, showing that, like $\beta(g)$, 
$\gamma(g_{xy})\simeq$ln$(g_{xy})$ deep into the metallic phase.  Thus, we have found that the {\it full} $dc$ conductivity tensor 
reflects the strong coupling nature of the ``Mott'' QCP, attesting to its underlying non-perturbative origin in Mott-like (strong scattering) 
physics.

\begin{figure}
\includegraphics[width=1.0\columnwidth , height= 
1.0\columnwidth]{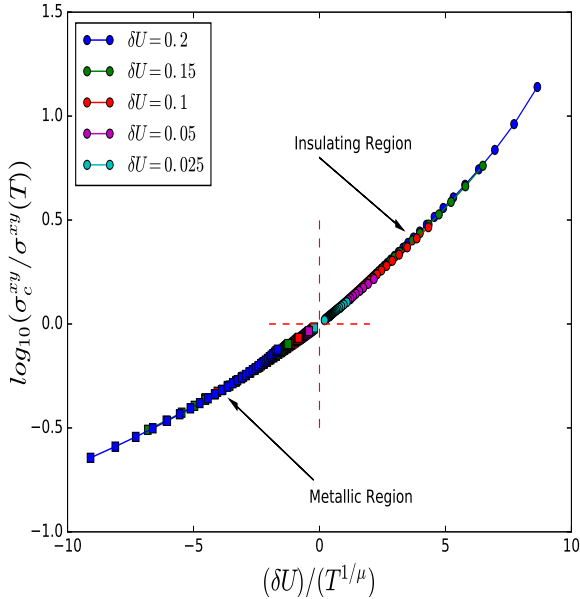} 
\caption{(Color online) $log_{10}(\frac{\sigma^{xy}_{c}}{\sigma^{xy}(T)})$  vs $(\delta U)/ T^{\frac{1}{\mu}}$, where $\delta U = U-U_c$.} 
\label{fig:fig5}
\end{figure}

That $\gamma(g_{xy})\simeq$ ln$g_{xy}$ holds on both sides of the MIT implies that its two branches must display ``mirror symmetry'' over an 
extended range of $g_{xy}$.  In Fig.\ref{fig:fig3}, left panel, we indeed see that magneto-transport around the QCP exhibits well-developed 
reflection symmetry (bounded by dashed vertical lines),  It is also manifest in the right panel of Fig.\ref{fig:fig3}, where 
$\sigma^{xy}_c/\sigma^{xy}(\delta U)=\sigma^{xy}(-\delta U)/\sigma^{xy}_{c}$; $i.e$, they are mapped onto each other under reflection 
around $U_{c}$, precisely as found earlier for the $dc$ resistivity.  As a final check, we also show (see Fig.~\ref{fig:fig5}) that 
log $(\sigma_{c}^{xy}/\sigma^{xy}(T)$ is a universal function of the ``scaling variable'' $\delta U/T^{1/\mu}$. Thus, our study explicitly shows the novel quantum criticality in 
magneto-transport at the ``Mott'' QCP (associated with a {\it continuous} Mott transition) in the FKM at strong coupling.

  In an Anderson model framework, scaling of $\sigma_{xy}$ is long known~\cite{abrahams}.  Our findings are very distinct from expectations for an Anderson-like transition: observe that we find $T^{xy}_{0}(\delta U)\simeq c_{xy}|\delta U|^{\mu}$ (in left panel of Fig.~\ref{fig:fig4}) with $\mu\simeq 0.75=3/4$ (in inset) on both sides of $U_{c}$, as required for genuinely quantum 
 critical behaviour.  This strongly contrasts with the $T_{0}^{xx}(\delta U)\simeq c|\delta U|^{z\nu}$ with $z\nu=1.32\simeq 4/3$ found for 
 the $dc$ resistivity~\cite{qcmott}.  Further, in the right panel of Fig.\ref{fig:fig4}, we also show that 
 $\sigma_{xy}=\sigma_{0,xy}(U_{c}-U)^{\mu'}$ with $\mu'=1.8$ (in inset), quite distinct from $\nu\simeq 4/3$ found earlier 
 for $\sigma_{xx}(U)$.  
 
 Along with our finding of $\sigma_{xx}(T)\simeq T$ and $\sigma_{xy}(T)\simeq T^{1.2}$ at the MIT, these findings have very interesting 
 consequences: $(i)$ the Hall constant is critical at the MIT.
We find $R_{H}^{-1}\simeq \sigma_{xx}^{2}/\sigma_{xy}\simeq (U_{c}-U)^{0.8}$, whereas $R_{H}$ is 
non-critical~\cite{abrahams} at the Anderson MIT, 
$(ii)$ $R_{H}$ is also strongly $T$-dependent and divergent at the MIT, varying like $R_{H}(T)\simeq T^{-0.8}$,
 whereas $R_{H}\simeq (nec)^{-1}$ in an Anderson disorder model.  Concomitantly, the Hall {\it angle} also exhibits anomalous 
behaviour: $(iii)$ tan$\theta_{H}(T)\simeq T^{0.2}$ and tan$\theta_{H}(U)\simeq (U_{c}-U)^{1/2}$ in the quantum critical region.
\begin{figure}
\includegraphics[width=1.\columnwidth , height= 
1.\columnwidth]{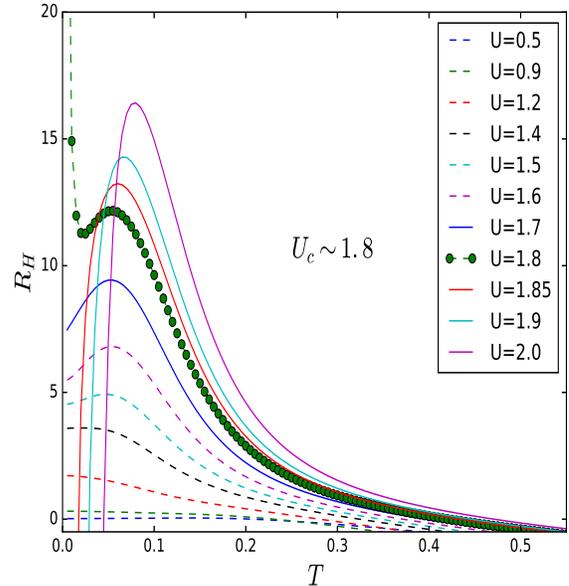} 
\caption{(Color online) Hall co-efficient $R_H$ as a function of temperature T for different U.}
\label{fig:fig6}
\end{figure}
   Our results are distinct from expectations from a Landau FL and Anderson-MIT views.  At an Anderson 
MIT~\cite{abrahams}, $R_{H}=(nec)^{-1}$ is $T$-independent and 
   non-critical at the MIT.  In the metallic phase, use of semiclassical
ideas dictates that {\it both} $\beta(g)$ and $\gamma(g_{xy})$ scale like $(d-2)-A/g$, and the quantum correction to the Hall conductance is 
twice as big as for the Ohmic conductance.  The stringent
assumption under which this holds is that the inverse Hall constant (related to $h(L)=L^{d-2}/R_{H}B$ in Abrahams {\it et al.}) scales 
{\it classically} like $h(L)\simeq L^{d-2}$ for small $B$ (large $h$).  
It is precisely this assumption that breaks down when one considers the Mott MIT, where $R_{H}$ 
{\it is} critical at the MIT (see above).  They are also different from expectations in a correlated LFL: a strongly $T$-dependent $R_{H}$ close to a Mott MIT in a Hubbard model framework is long known~\cite{hrk-1994}.  However, in a DMFT framework, $R_{H}(T)$ exhibits a recovery of correlated Landau-Fermi liquid behaviour below a low-$T$ lattice coherence scale.  Moreover, the MIT there
is a first-order transition.  In the FKM, the metallic state remains bad-metallic and incoherent down to lowest $T$, and the MIT is continuous.  The Mott-like character of the associated QCP is revealed by the 
observation of $\beta(g) \simeq$ log$(g)$ and $\gamma(g_{xy}) \simeq$ log$(g_{xy})$. 

In Fig.\ref{fig:fig6}, we show $R_{H}(U,T)$ versus Temperature (T).  
Both are indeed
 markedly $T$-dependent.  For an Anderson MIT, $R_{H}$ would be non-critical.  In a LFL metal, one expects
$\sigma_{xx}(T)=1/\rho_{dc}(T)=AT^{2}$, while $\sigma_{xy}(T)\simeq T^{-4}$ at low $T$.  In that case, we end up with a $T$-independent $R_{H}$ and cot$\theta_{H}(T)=cT^{2}$.  This is the expected behaviour for a LFL, where a single relaxation rate governs the $T$-dependent relaxation of longitudinal and Hall currents.
 Very different $T$-dependences we find here testify to the breakdown of this
intimate link between LFL quasi-particles and this conventional behaviour, and
that the results we find are direct consequences of the destruction of LFL quasi-particles at strong coupling.  They render semiclassical Boltzmann arguments (based on validity of   $k_{F}l>>1$) inapplicable at the outset.
 
We now turn to experiments to investigate how our theory stands this stringent test.  Recent 
work on NbN~\cite{madhavi} most clearly reveals ill-understood signatures of localization incompatible with weak localization predictions.
\begin{figure}
\includegraphics[width=1.0\columnwidth , height= 
1.375\columnwidth]{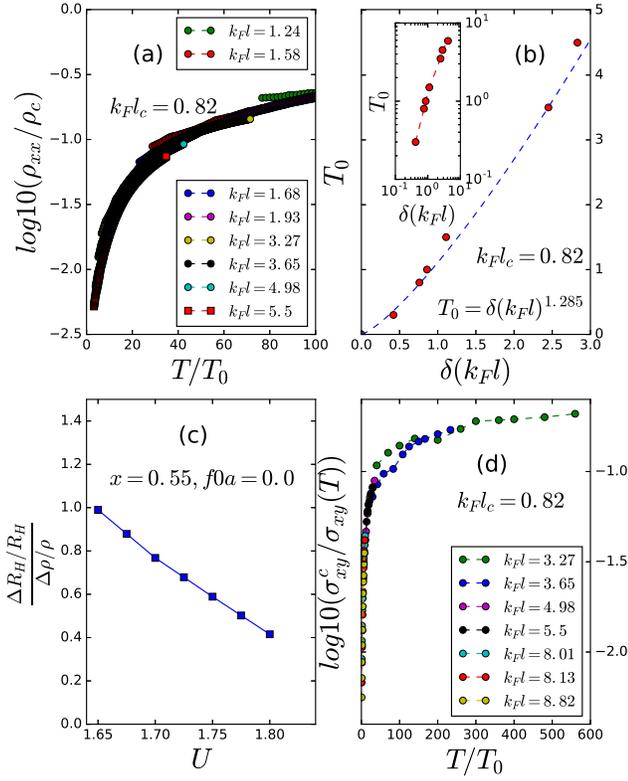} 
\caption{(Color online) (a). Resistivity data from Chand {\it et al.}~\cite{madhavi}, replotted as log$(\rho_{xx}(T)/\rho_{c})$ versus 
$T/T_{0}$ with $T_{0}(\delta k_Fl)\simeq |\delta k_F l|^{1.3}$ in Panel (b), in excellent accord with theory~\cite{qcmott}.  In Panel (c), we show that
the theoretical ratio $\frac{\Delta R_{H}/R_{H}}{\Delta\rho/\rho}$ is in the range of $0.5-0.7$ near the Mott QCP, again in good qualitative 
accord with the value of $0.69$ from Hall data~\cite{madhavi}.  In Panel (d), we show clear scaling of the {\it experimentally} 
extracted log$(\sigma_{xy}^{(c)}/\sigma_{xy}(T))$ in very good accord with theory for the same sample set used for Panel (a). The $\sigma_{xy}(T)$ is constructed from the experimental dc resistivity and Hall constant ($R_H$). The $R_H$ at the critical $k_Fl$ is calculated from extrapolation of the experimental Hall constant ($R_H$) down to $(k_{F}l)_{c}=0.82$, as shown in Fig.~\ref{fig:fig2SI}}
\label{fig:fig7}
\end{figure}
\begin{figure}
\includegraphics[width=1.\columnwidth , height=
1.\columnwidth]{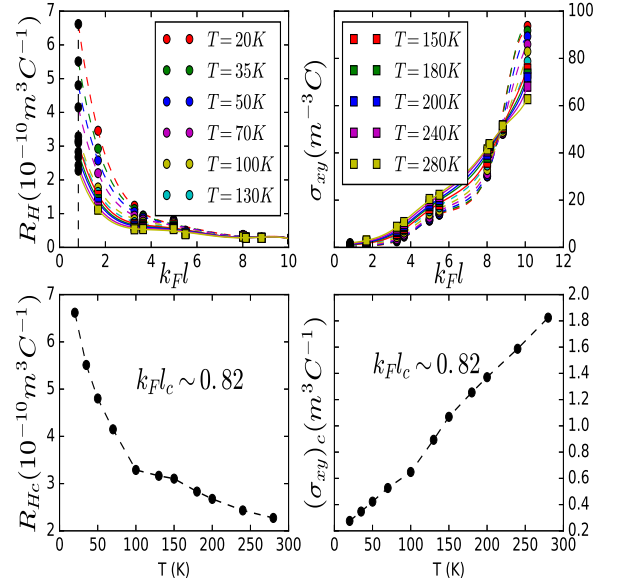}
\caption{(Color online)(a) In top-left panel, Hall constant $R_H$ as a function of $k_Fl$ for various temperature T, black circles are the values of $R_H$ after extrapolating the curves upto critical $k_Fl=0.82$ (b) In top-right panel, Hall conductivity $\sigma_{xy}$ calculated from the Hall constant ($R_H$) and dc resistivity ($\rho_{xx}$) as a function of $k_Fl$ for various temperature T (c) In bottom-left panel, Hall constant $R_H$ as a function temperature (T) at the critical $k_Fl=0.82$ calculated from the extrapolation, (d) In bottom-right panel, Hall conductivity $\sigma_{xy}$ as a function temperature (T) at the critical $k_Fl=0.82$  }
\label{fig:fig2SI}
\end{figure}
In NbN, the effect of intentional charge disorder is to cause a random variation in the local atomic potential, which increases 
as $k_{F}l$ is reduced by increasing the disorder level.  Following Freericks {\it et al.}~\cite{freericks}, we posit that the FKM is a suitable effective model for materials like Ta$_{x}$N and $Nb_xN$, where carriers interact locally with randomly distributed charge disorder.
 We have reanalyzed Chand {\it et al.}'s data on NbN in light of the above results 
to test how our strong coupling view performs relative to data.  To make meaningful contact with data on NbN,
we make a reasonable assumption that increasing $U/t$ in the FKM corresponds to {\it decreasing} $k_{F}l$, 
since the scattering strength should increase with $U/t$, reducing $k_{F}l$ to $O(1)$~\cite{madhavi} near the MIT.
We find: as shown in Fig.~\ref{fig:fig7}(a), that $(i)$ log$(\rho_{xx}(T)/\rho_{c})$ on the (bad) metallic side scales with $T/T_{0}(k_{F}l)$ 
exactly as predicted by our theory~\cite{qcmott}.  Further, the data analysis shows (Fig.~\ref{fig:fig7}(b)) that 
$T_{0}(k_{F}l)\simeq (k_{f}l-(k_{f}l)_{c})^{z\nu}$ with $z\nu\simeq 1.3$, again in excellent accord with theory if we identify 
decreasing $k_{F}l$ with increasing $U$ in our model.  $(ii)$ interestingly, our $\rho_{xx}(T),R_{H}(T)$ results reproduce the detailed 
$T$-dependence seen in data~\cite{madhavi} with only one adjustable parameter ($U$).  $(iii)$ even more remarkably, we find that 
$(\Delta R_{H}/R_{H})/(\Delta\rho_{xx}/\rho_{xx}$, shown in Fig.~\ref{fig:fig7}(c), achieves values between 0.5 and 0.7 close to the MIT 
(between $1.5\leq U \leq 1.9$) in our model, in very good accord with 0.69 extracted in experiment.  Finally, in Fig.~\ref{fig:fig7}(d),
we uncover quantum critical scaling in $1/\sigma_{xy}(T)$ as a function of $k_{F}l$ from data on NbN, which is expected in our model, since
both $\sigma_{xx},\sigma_{xy}$ 
exhibit such novel scaling behaviour.  Since $R_{H}$ is difficult to extract reliably in very bad-metallic samples (with $k_{F}l<3.0$) close to the MIT, we resorted to a careful extrapolation of the Hall conductivity ($\sigma_{xy}$) and Hall constant ($R_{H}$) to smaller values of $k_{F}l$.

In Fig.~\ref{fig:fig2SI}, we show the results of a careful fitting of the experimental data down to
$k_{F}l\simeq O(1)$ (in fact, the critical $(k_{F}l)_{c}$ is now consistent with $0.82$, which is the critical value for the longitudinal dc conductivity).  Using these extrapolated fits to the dc conductivity tensor as a function of $k_{F}l$, we constructed Fig.~\ref{fig:fig7}(d) in the main text.  This makes our analysis consistent with a single $(k_{F}l)_{c}\simeq 0.82$ for {\it both} $\sigma_{xx}(k_{F}l,T)$ 
and $\sigma_{xy}(k_{F}l,T)$.

%\begin{figure}
%\includegraphics[width=1.\columnwidth , height= 
%0.53\columnwidth]{fig7a.eps} 
%\caption{(Color online) }
%\label{fig:fig6}
%\end{figure}

Taken together, earlier results of Chand {\it et al.}~\cite{madhavi}, now suitably reanalyzed in light of our CDMFT results, receive 
comprehensive explication within a ``strong localization'' view adopted here, lending substantial support to the view that the novel findings 
in NbN are representative of strong scattering effects near a continuous MIT, and involve microscopic processes beyond perturbative-in-$(1/k_{F}l)$ approaches.

   Thus, to conclude, we have presented clear evidence of novel quantum critical behaviour in magneto-transport near a continuous MIT by a 
   careful scaling analysis of CDMFT results for the off-diagonal conductivity for the FKM in the strong localization limit.  We find that
the loss of the quasiparticle pole structure at strong coupling ($k_{F}l\simeq 1$) leads to a rather distinct ``Mott''-like quantum
criticality, necessitating substantial modification of the quasiclassical Drude-Boltzmann transport schemes to study (magneto)-transport.  
The resulting quantum criticality we find is closer to that expected from the opposite limit of strong localization based on a real-space
locator expansion~\cite{anderson1958,sudip1997}, as manifested in $\gamma(g_{xy})\simeq$ ln$(g_{xy})$.  Comprehensive and very good explication 
of recent data on NbN lend substantial experimental support to this Mott-like view.  We suggest that strongly disordered
electronic systems that show a bad-metallic resistivity and sizeable $T$-dependent Hall constant would be promising candidates to unearth 
such novel quantum-critical magneto-transport at a continuous MIT.  Finally, the similarity of QC scaling in resistivity
in earlier work~\cite{qcmott} to the Mott QC scaling in the Hubbard model~\cite{terletska} above the finite-$T$ critical end-point 
suggests that related features discussed above may also manifest in wider classes of strongly correlated Mott materials.

\end{document}